\documentstyle[multicol,aps,prl,epsf,psfig,draft]{revtex}
\def\be{\begin{equation}}
\def\ee{\end{equation}}
\def\bea{\begin{eqnarray}}
\def\eea{\end{eqnarray}}
\def\ba{\begin{array}}
\def\ea{\end{array}}
\def\bc{\begin{center}}
\def\ec{\end{center}}
\def\bi{\bibitem}

\begin{document}
\title{ Testing a hypothesis for the evolution of sex}
\author{ Bora \"Or\c cal$^{1}$, Erkan T\"uzel $^{1,2}$,
Volkan Sevim$^{1}$, Naeem Jan$^3$, and Ay{\c s}e Erzan$^{1,4}$}
\address{$1$ Department of Physics, Faculty of  Sciences
and Letters\\
Istanbul Technical University, Maslak 80626, Istanbul, Turkey\\
$^2$ Department of Physics, I\c s\i k University, Maslak, 80670,
Istanbul,
Turkey\\
$^3$ Department of Physics, St. Francis Xavier
University, Antigonish, Canada\\
$^4$  G\"ursey Institute, P. O. Box 6, \c
Cengelk\"oy 81220, Istanbul, Turkey}
\date{\today}
\maketitle
\begin{abstract}
An asexual set of primitive bacteria is simulated with a
bit-string Penna model with a Fermi function for survival.
A recent hypothesis by Jan, Stauffer and Moseley on the evolution of sex
from asexual cells as a strategy for trying to escape the effects
of deleterious mutations is checked. This strategy is found to
provide a successful scenario for the evolution of a stable
macroscopic sexual population.

%PACS 87.23.Kg, 5.40.-a
\end{abstract}
%\newpage
%%%%%%%%%%%%%%%%%%%%%%%%%%%%%%%%%%%%%%%%
%%%%%%%%%%
\begin{multicols}{2}
\section{Introduction}
Asexual reproduction is the complete and faithful
expression of all the genes of the {\it mother} cell in the {\it daughter}
cells. It is efficient and straightforward. Why then did {\it sex} evolve?
~\cite{Hurst,Smith} Jan, Stauffer and Moseley have proposed~\cite{Jan}
a small environment (small population) with asexual one-celled organisms,
(which we will denote as ``bacteria'' from now on) in which deleterious
mutations are driving some into extinction. It is postulated that these
soon to be extinct bacteria may indulge in sexual reproduction as a last
resort, to give rise to offspring that are better suited to the
environment. It is the purpose of this paper to provide a partial test
of whether this strategy will increase
their chances of propagation in the immediate evolutionary game.

%%changes%%
The genome of each bacterium is represented
by a double bit-string~\cite{Penna}. We use the term {\it wildtype} to
represent
the bit-string that is best adapted (ideal) to the environment. Asexual
bacteria have two identical bit-strings (they are ``haploid") whereas
sexual types are ``diploid," i.e., they evolve a pair of bitstrings that
may be quite distinct. In this very rudimentary model, each bacterium may
be regarded as having only one chromosome - so that the ``law of
independent assortment"~\cite{Pai}
does not hold here - all the genes are linked.
%%

%%changes%%
The salutary effect of sexual reproduction comes from the important
assumption that we make, namely, that {\it deleterious mutations are
recessive}.
%%

%%changes%
In this paper ``sexual reproduction'' will mean a process whereby {\it i)}
a number $n$ of germ cells are formed from each parent cell (meiosis).
These germ cells contain half the amount of genetic material present in
the parent cell, i.e., only one bit-string. {\it ii)} New individuals are
formed by pairing germ cells (i.e., single bit-strings) from two 
parent cells. In this way, the parents are replaced by $n$ offspring. We have
confined ourselves to $n=1$ or $n=2$. There is no differentiation between
the sexes.

We investigate several alternative scenarios for the reproductive rules.
In the first, (called Model I below)
 individuals undergo a mutation
which enables them to engage in sex as an extreme survival measure.
Thier offspring subsequently reproduce asexually (by mitosis or
``simple fission") under less harsh conditions.  In this scenario, we find that
the asexual population becomes extinct, and the ``sexual types" eventually
win over the population.

%%changes%%
In the second scenario, the descendants of sexual types always reproduce
sexually. To safeguard against the number of sexuals dropping too
drastically, we first took  $n$, the number of offspring, to be two,
but the recent converts were still only allowed to mate amongst each other.
This is called Model II below, and gave rise to a macroscopic
sexual population. Then, we investigated what happens if the recent
converts (all of which are in the danger zone, facing extinction) did not
just mate amongst each other, but were allowed to pick mates from the
better adapted sexual population at large.  This was tried both for
the case of $n=2$ (Model III) and $n=1$ (Model IV).
Finally, getting bolder, we tried
the case where $n=1$, and moreover the recent converts are only allowed to
mate amongst each other (Model V). We found that in all of these cases,
the steady state population comprised a finite fraction of sexual types.
with the fraction being dependent on the
number $n$ of off-spring, and on the rules according to which the
individuals may choose their mates.
Thus in these
models with varying degree of bias against the sexual population,
we have
found that the survival of the sexual population is rather robust.
%%

%%changes%%
In all the models we have adopted the convention~\cite{Redfield} that the
total population is kept fixed.  This is accomplished by duplicating a
sufficient number of asexual bacteria in each cycle to make up for the
attrition due to deaths or to sexual reproduction with $n=1$. The
efficacy of a particular mode of reproduction is measured by the long
term representation in the population of the types engaging in that
particular mode of reproduction, i.e., sexual v.s. asexual.

The paper is organized as follows.  In section 2 we define our
models.
We give enough details to enable further
simulations and encourage independent checks of our results.
In section 3 we present our findings from the simulations.
In section 4 we state our conclusions.
%%%%%%%%%%%%%%%%%%%%%%%%%%%%%%%%%%%%%%%%
%%%%%%%%%
\section{Models for the Evolution of Sex in One Celled Individuals}

We represent the genetic code of each one-celled
individual with a bit-string of ``0"s and ``1"s, after the
Penna model~\cite{Penna}.
For asexual, haploid, cells, we have two 16-bit strings that are identical
copies of each other. For the sexual cells, we have two 16-bit
strings(``gametes'') which are allowed to be different, i.e., the
individuals are now diploids.  We use the bit
defining the ``sign", to specify whether the individual is sexual
or asexual - negative (1) indicating sexual and positive (0)
indicating asexual.

\subsection{Asexual steady state}

We start with a set of $N$ initially identical asexual individuals,
all identical to the {\em wildtype}, i.e., all $0$'s.
The probability of a mutation hitting any individual is $\Gamma
= 1/N$ at any step, and it is implemented by
scanning all the individuals in the population, and mutating each
individual with a probability of $1/N$.  Clearly
there may be any number of mutated individuals at any one
time step, the number fluctuating around unity.
Mutations are
defined as the operation of addition modulo 2, applied to a randomly
chosen bit in the string, except the sign bit. Alteration of the sex
gene takes place only under special conditions, namely the threat of
extinction due to too many deleterious mutations.
For the asexual individuals, mutation of any
one of the bits affects both strings.

The probability of survival, for individuals who have
experienced $m$ mutations, is given by a Fermi-like
distribution~\cite{Thoms},
$P(m)$,
\be P(m) = { 1  \over {\exp[\beta (m - \mu)] + 1}} \;\;\;,
\label{Fermi} \ee
where $m=0,1,\ldots L$,
for a bit string of length $L$.
For large $\beta$ (or ``low temperatures," in the language of
statistical mechanics), $P(m)$ behaves like a step function.
Individuals with $m>\mu$ die, those with $m<\mu$ survive, and those
with $m=\mu$ survive with a probability of $1/2$.

At each time step, all individuals are subjected to the fitness
criterion represented by this function - i.e., each survives with
probability $P(m)$, depending on the number of mutations it
has at the moment. (In the simulations we report below, we set
$\beta=10$ and $\mu=4$.)

The model defined so far clearly describes a random walk in one
dimension (the number of mutations),  with a sink at $m\ge \mu$ for large
$\beta$.  With $n_a(m,t)=0$ for $m<0$ and $m>L$, $n_a(m,t)$ obeys the set of
equations
\bea
{\partial n_a(m,t)\over \partial t} & = & \sum_{\delta=\pm 1} 
[ T_{m+\delta,m} n_a(m+\delta,t) \cr
& - & T_{m,m+\delta} n_a(m,t) ] 
-[1-P(m)] n_a(m,t). \label{master}
\eea
There is a drift towards larger values of $m$,
since the stepping rates $T_{m,m+1}=\Gamma (L-m)/L$ and
$T_{m,m-1}= \Gamma m/L$. For $L>2\mu$, as is the case here,
$T_{m,m-1} < T_{m,m+1}$. The population would decay exponentially
to zero, if it were not replenished by reproduction.
We keep the total population constant, as in the Redfield
model~\cite{Redfield}, by
making up for the deficit in the population
after all the bacteria have been either found fit for survival or
killed off according to the survival probability in
Eq.~(\ref{Fermi}).

An early stage of evolution, (before ``sex is introduced'') can be
modeled by purely asexual reproduction.
We make up for the decrease $\delta N$ in the population by
randomly selecting $\delta N$ surviving bacteria and replicating
them once.
This corresponds to adding a source term proportional to
$[N-\sum_{m^\prime} n_a(m^\prime,t)]n_a(m,t)$ to the RHS of the
master equation~(\ref{master})
for the distribution of our asexual population, $n_a(m,t)$.
Running through many such cycles, one finds that the population
settles down to a steady state state distribution $n_a(m)$.

It should be noted that in our model, sexual reproduction at best keeps
the population constant, as outlined below, and therefore it is always
asexual reproduction that augments the population to make up for the
deficit, even after sex has been turned on.

%%changes%%
\subsection{Sexual types and sexual reproduction}
In this paper, sexual types will be distinguished from asexuals by two
features:  {\it i)} they are diploids and
{\it ii)} they may reproduce sexually.  We now specify what these mean.
{\it i)} Once the
``sex gene" is turned on,
we allow the two bit-strings of sexual types
to be different. This makes
room for greater genetic
variety. Moreover, since we take  each gene to have
an equal and independent
probability to be mutated, for diploids, the probability of any
gene to be mutated is halved in comparison  to the haploid types.
All this gives the sexual individuals a greater chance of survival than
the haploid asexuals~\cite{Martins}.

{\it ii)} Sexuals may  engage in
``sexual reproduction."
We have considered two variants of sexual reproduction (see Fig. 1),
depending upon the number of germ cells and subsequent number of
offspring.
\begin{description}
\item[\hspace{3pt} S1 \hspace{2pt}]
Two sexual
bacteria mate to give rise to one sexual offspring.
In this case, each parent cell  undergoes meiosis to produce one germ
cell, which posesses one of the bit strings (randomly chosen) of the
parent cell.
The germ cells coming from  the two parents merge to form a "daughter."
Thus, the  ``daughter'' has
a pair of bit-strings (``gametes'') each coming from one of the
parents, randomly selected from the four such pairs that one may
form out of the gametes  of the parents. In this definition, the
population is reduced by one each time an act of sexual reproduction
(``mating")  takes place.
\item[\hspace{3pt} S2 \hspace{2pt}] Two sexual
bacteria mate to give rise to a pair of offspring.
In this case, each parent gives rise to two germ cells, which combine to
form two daughters.
The gametes of the parents (say $Aa$
and $Bb$) can be shared between the offspring in two different ways, i.e.,
$(AB, ab)$ or $(Ab, aB)$.
The
population stays constant.
\end{description}
%%

%%changes%
\subsection{The Dominant String}
Since the sexual individual has two different gametes,
or bit-strings, we have used the concept of
the ``expressed'' or ``dominant" string, to compute the
survival probability.

We assume at the outset that deleterious mutations are recessive. The
way we have implemented this in practice is as follows.
Once a sexual offspring comes into being,
we form  the ``dominant''or ``expressed" string, by comparing
each bit with the wildtype and actually
exchanging bits between the strings
to make the ``expressed'' string as close to the wildtype as
possible. Clearly, this ``expressed" string has fewer
deleterious mutations than either of the two strings coming from the
germ cells making up this individual, and similarly,
the other string is now worse
off (has more deleterious mutations).
Once this reshuffling has taken place and the ``dominant string" has
been formed, any further mutations that happen to hit this string are
considered dominant, and $m$ is always computed by counting
the deleterious mutations on this ``expressed" string.
The germ cells of this parent will
now pass on these reshuffled gametes, possibly further modified
by subsequent mutations, to their offspring.

It should be strongly noted that in exchanging bits between the
bit-strings in this deterministic way we have incorporated a
feature into our model which is called a ``meiotic 
drive"~\cite{Smith}, occuring rarely in nature.  
Although in neglecting to
bring into play dominance/recessiveness in subsequent mutations, i.e., 
after the ``dominant string" has been formed, we have an element 
which counteracts the meiotic drive to a certain extent, the 
way in which dominance is handled here is not very realistic.
This will be further discussed in the last section.
%%

%%changes%%
\subsection{Conversion to sex}

Faced with a crisis situation, i.e., th number of deleterious
mutations $m$ becoming too large and
threatening survival, we assume that the
bacteria engage in sexual reproduction.
For all the different models that we have considered,
once the asexual steady state is achieved, we allow the sex gene to
be ``turned on'' for the least fit members of the population.
In any pass through the population,
if those individuals
that are in  the tail of the distribution  (i.e. those with $m
\geq \mu$ mutations) survive, then they are
turned sexual by deterministically and irreversibly switching their sign
bits to one.

The next two steps make up the reproductive cycle:
Once their sex bit is turned on, these individuals will
be ``sexually active'' and mate with another
sexual individual. In the last step of the reproductive
cycle,
the population is allowed
to grow back to its fixed value.

We have considered several {\bf Models
(I-V)} which differ from each other in the details of the reproductive
cycle: whether and when the sexual types
reproduce sexually or asexually, the number of offspring and
the choice of mate.
We define these models in detail
below. We then go on to give a synopsis of all the steps involved
in one complete pass, indicating how each step differs from model to
model.

\begin{description}
\item[Model I]  Here sex is only used by sexuals in
time of crisis.
Sexually active individuals ($m\ge
\mu$) {\bf mate amongst each other} according to rule {\bf S1} and beget one
offspring. Once out of the ``danger zone" (i.e., for  $m < \mu$), bacteria
reproduce asexually, regardless of whether they are sexual or asexual
types.  Thus, in the last step of the reproductive cycle, if a sexual type
is picked at random as a candidate for reproduction, it simply 
undergoes mitosis, as would an asexual type.
\end{description}
In Models II-V, sexual individuals are only
allowed to reproduce sexually. To preserve the symmetry with
Model I, however, we have allowed all individuals to be sampled
in the last step of the reproductive cycle. If the random sampling yields
a sexual individual, it has to reproduce sexually according to the
procedure specified below for that model.  If the random choice yields an
asexual type, then it undergoes mitosis.
\begin{description}
\item[Model II]
Sexually active individuals ($m\ge \mu$) {\bf mate amongst each other}
according
to rule {\bf S2}, begetting two offspring. In the last step of the
reproductive cycle, if a randomly picked candidate for reproduction
happens to be sexual, it is mated with another randomly picked sexual
and reproduces according to {\bf S2}, leaving the population constant.
\item[Model III]
Sexually active individuals ($m\ge \mu$) pick a {\bf mate from the sexual
population at large}, and mate with it
according
to rule {\bf S2}, begetting two offspring. In the last step of the
reproductive cycle, if a randomly picked candidate for reproduction
happens to be sexual, it is mated with another randomly picked sexual
and reproduces according to {\bf S2}, leaving the population constant.
\item[Model IV]
Sexually active individuals ($m\ge \mu$) pick a {\bf mate from the sexual
population at large}, and mate with it
according
to rule {\bf S1}, begetting one  offspring. In the last step of the
reproductive cycle, if a randomly picked candidate for reproduction
happens to be sexual, it is mated with another randomly picked sexual
and reproduces according to {\bf S1}, reducing the population by one.
\item[Model V]
Sexually active individuals ($m\ge
\mu$) {\bf mate amongst each other} according to rule {\bf S1} and beget one
offspring, as in Model I.
However, unlike Model I, in the last step of the reproductive cycle, if a
sexual type
is picked at random as a candidate for reproduction, it mates with another
randomly picked sexual and begets one offspring according to rule {\bf S1},
thereby reducing the population by one.
\end{description}

In summary,
in Models I, II and V, individuals turned sexually
active in the face of extinction, with $m\ge \mu$, mate among each other,
while in Models III and IV,  they are allowed to pick their mates from
among the sexual population at large, thus having a chance to mate with
$m<\mu$ individuals closer to the wildtype.  On the other hand, while in
Models II and III, sexual reproduction does not reduce the number of
sexual types (rule {\bf S2}), in Models I, IV and V, it reduces it by one
(rule {\bf S1}) everytime it occurs.

As we will see in the next section,
these choices lead to  different results.  In Model I, the
asexual population grows extinct and the sexuals completely win over.
For Models II-V, we find that the steady state comprises a finite fraction
of sexuals.

%%changes%%
\subsection{The kinetics including sex}
A complete pass  now consists
of the following steps:

\begin{enumerate}
\item {\bf Mutation and Decimation}
Each individual is  subjected to the possibility of
a mutation at the rate of $\Gamma$, independently of whether it is
sexual or asexual.

For an asexual individual,
one proceeds as described in subsection II.A, and the individual
either survives with a probability $P(m)$ or is killed off
with probability $1-P(m)$.

If a sexual individual is hit by a mutation, one
of the two
bit strings is chosen with probability $1/2$; then one bit on this string
is chosen randomly (with probability $1/L$) and mutated.
The number of mutations $m$, and subsequently
the survival probability $P(m)$, are computed with
respect to the ``dominant'' string, as described in Section II.C.

\item {\bf Conversion} Of the surviving asexuals those
with $m\ge \mu$ are turned into sexuals, and tagged ``sexually active."
If a sexual individual with $m\ge \mu$ survives in a
given pass,
then  it is  also tagged ``sexually active.''

\item {\bf Reproduction 1} At the end of one
complete cycle of mutations, decimation or conversion,
all the ``sexually active''
bacteria are made to reproduce according to the following rules:
\begin{description}
\item[Model I] We randomly form  pairs of all  ``sexually active" bacteria
($m\ge \mu$). They  reproduce
according to {\bf S1}, each pair begetting one offspring.
\item[Model II] All  ``sexually active" bacteria
are paired as above,  and reproduce according
to {\bf S2}, each pair begetting two offspring.
\item[Model III] Each ``sexually active" bacterium ($m\ge \mu$) picks
a mate at random, from the sexual population at large.  They mate
according to rule {\bf S2}.
\item[Model IV] Each ``sexually active" bacterium ($m\ge \mu$) picks
a mate at random, from among the sexual population at large.  They mate
according to rule {\bf S1}.
\item[Model V] All ``sexually active" bacteria are
randomly paired among each other and reproduce according to
rule {\bf S1},  as for Model I above.
\end{description}
The offspring are tagged ``sexually inactive," so that they are not
to be  mated
in this reproductive step.
In Models I, II, IV and V,
if the number of sexually active bacteria is odd, so that there is an odd
guy out after the random pairing,
it is still ``active'' but will have to
await the next cycle to see if it gets a mate.

\item {\bf Reproduction 2} At this (second) step of the reproductive
cycle we allow the population to grow back to $N$ by means of the following
rules:
\begin{description}
\item[Model I]  Out of the surviving population, we randomly
pick $\delta N$ individuals and make them reproduce asexually
(i.e., simply replicate them), regardless of whether they are sexual
or asexual.
\item[Models II, III]
Out of the surviving population, we start to pick
out individuals at random.
If the chosen individual is asexual, then
it reproduces asexually by replication, thereby augmenting the
(asexual) population by one.
If the individual is sexual, then another individual
is picked out of the sexual population at large, and they
reproduce sexually  according to rule {\bf S2},
which leaves the population unchanged.
\item[Models IV, V]
Out of the surviving population, we start to pick
out individuals at random.  If the chosen individual is asexual, then
it reproduces asexually by replication.  If the individual is sexual,
then it mates with another individual out of the
sexual population at large, and they reproduce
sexually  according to rule {\bf S1}, which means that
the (sexual) population is diminished by one.
\end{description}
We proceed in this manner
until sufficiently many individuals have been added to the population
so that  the total has been restored to $N$.
\end{enumerate}
It can be seen  that in those cases (Models II-V)
where the sexual bacteria
are not allowed to regress and reproduce asexually,
the sexuals have a disadvantage in the number of
offspring per parent, and they will owe their survival to their
strategic advantage of being able to improve their fitness due
to sexual reproduction. To recapitulate,
in Models II and III, a pair is  allowed to
have two offspring, a feature which gives the sexuals less of a
disadvantage than in Models IV and V. The feature which distinguishes
Model II from III (and Model V from IV) is that in the former,
those bacteria turning to sex at the edge of extinction are only allowed
to mate among themselves, whereas in the latter, they are allowed to
pick their mates from among the sexual population at large.  One would
naively expect that the conditions are more stringent for Model II
(and Model V) than they are for Model III and IV, since the former have a
larger variety of fitter individuals to mate with.  A surprise awaits us
in the next section.

%%%%%%%%%%%%%%%%%%%%%%%%%%%%%%%%%%%%%%%%
%%%%%%
\section{Sex Succeeds}

We performed the simulations for the above models on a fixed population
of $N=256$, for 16-bit strings.
The equations for the evolution of the sexual and asexual populations,
$n_a(m)$ and $n_s(m)$,
are nonlinear in these quantities.
Therefore we checked in every case that there
was, at least typically,
no ergodicity breaking, and no
periodic or strange attractors
for the dynamics, by performing 100 different runs
for each set of rules.  The results which we quote in the tables are
averaged over 100 runs.  The fluctuations are still relatively large,
with a relative error estimate based on one standard deviation typically
being  $6 \%$ for the bar graphs shown in Figs. (2-7).

Before sex is turned on, we find that the asexual population reaches
a steady state distribution with respect to the number of mutations.
The
average distribution for the asexual steady state
is given in Table I and Fig. 2.
(In this and the subsequent bar graphs, for each $m$ we report
percentages of the
total population, to make it easier to grasp the figures.)
We see a population
that is fixed at $N=256$ but there are almost no `wildtype'
bit-strings; on the other hand the graph is peaked at $m=3$, which
is the ``minimally stable'' value of $m$.
Note that this distribution is similar to that seen for the
self-organized critical state of the sandpile~\cite{Bak}, where $m$
plays the same role as the units of sand at a particular site.

Once sex is turned on in Model I, it takes a time roughly proportional
to the size of the population for asexual individuals to become extinct.
Our results for the relaxation time,
averaged over 100 runs, still show quite a bit of fluctuation,
but are  approximately  $10^3$, $2\times 10^3, \ldots 2^3 \times 10^3$,
for $N=32, 64, 128$ and 256.

The distribution over $m$, of the asexual and sexual types, have been
computed as ensemble averages over 100 copies of the system, with
the population fixed at $N=256$ in each case.  The initial state is
always taken with all individuals  identical to the wildtype.  Each system
evolves for 5000 generations and therefore surely reaches the asexual
steady state.  Then sex is turned on, and each system now evolves for
another thirty thousand generations.  Then the averages are taken over
the independent systems.

Within Model I, the sexual population reaches a steady state (see Fig.
3) still exhibiting a  peak at $m=3$; however, this peak is
slightly suppressed in comparison to its value in the asexual steady state,
whereas the population at $m=1$ is slightly augmented and there is a
nonvanishing population of wildtypes.  This
demonstrates that the sexual individuals are better capable of
eliminating deleterious mutations from their expressed genes.

The results for Model II are drastically different.  After sex is turned
on, one reaches a state of coexistence between the asexual and
sexual populations.  The asexual population has a distribution with
respect to the number of mutations  which is the same as in Fig.2,
whereas the distribution of sexual individuals has shifted markedly
towards lower values of $m$ as can be seen in  Fig.4.
For the sexuals, there is
a rather broad peak around
$m=1$, with an appreciable population of wildtypes.  The numbers for the
steady state populations of Models II-V are given in Table II, and
the total fraction of sexual and asexual populations are shown in the
pie charts in Fig. 8.

We see from Fig. 5 that the results for Model III are only
marginally different from those of Model II, but the difference is
in a direction we did not initially expect:  in all the $m$ values, the
percentages of the sexual population is slightly {\it higher} in
Model II than in Model III, and the total sexual population is a
also few
percentage points  higher in Model II (see Fig. 8).  This rather
small difference, which could be ascribed to a fluctuation, gets amplified
when one allows only one offspring per parent, as we do in Models IV and V.

Turning to Models IV and V (Figs. 6,7) we see that the feature of
producing {\it relatively} much better fit offspring (compared to both
parents), which we get when sex-out-of-desperation is constrained
to take place exclusively between  $m\ge \mu$ individuals (Model V)
again outweighs the advantage of being able, as in Model IV, for an
$m\ge \mu$ individual to be able to mate with a better fit partner
chosen from among the sexual population at large.  The total sexual
population in Model V is $5 \%$ larger than in Model IV.

%%%%%%%%%%%%%%%%%%%%%%%%%%%%%%%%%%%%%
\section{Conclusions}
%%changes%%
Our findings are consistent with
the hypothesis by
Jan et. al~\cite{Jan} that sex, practiced
as a last resort between individuals
on the verge of extinction, might give rise to a stable sexual population.
It remains to be investigated whether a finite rate of conversion of the
asexual population to sexual for {\em arbitrary} $m$, also leads to a
steady state sexual population, as found here.

It should also be noted, that ``meiotic parthenogenesis"(MP) is an
alternative
strategem whereby bacteria may escape the mortal effects of deleterious
mutations, without sexual reproduction~\cite{Smith,Bernardes}. 
This refers to the random
exchange of sub-sequences of genes  between the two bit-
strings (paired chromosomes) of a
diploid individual. In testing the Jan et al. hypothesis~\cite{Jan} 
we have not taken into account this rival strategy.

It is interesting to remark that in an alternative scenario~\cite{Martins2} 
a genetical catastrophe can eliminate an asexual, parthenogenetic
population,
while a sexual population can  survive.  
We have checked the mortality rates
for Model I (the most catastrophic for the asexual population), and found
that the model does not harbour any genetic catastrophes.  One might have
thought that its similarity to a ``sandpile model''~\cite{Bak} might give 
rise
to intermittently occuring mass deaths (avalanches), with a power
law distribution of casualties for large time scales, but this does
not turn out to be the case.  The number of deaths is typically small,
never exceding five for our population of $N=256$.

We would like to caution that, in 
the way we have implemented the formation of the "expressed
genotype," an element of ``meiotic drive" has actually crept into the model.
In forming the ``expressed string,"
genes (bits) are being exchanged between the two
bit-strings in a
way that is not even random, but highly purposeful.  In the subsequent
meiotic stage, this gives rise to two gametes one of which is much
closer and the other much farther from the wildtype than either of the
gametes of the parent as it was first formed.  This mechanism provides a
much stronger ``mixing" of the gene pool in this model
than afforded by
sexual reproduction plus the recessiveness of the deleterious mutations, 
and does not typically occur in nature~\cite{Smith}.
Further work is in
progress to remove this spurious effect.

We may finally conclude that our model incorporates a delicate balance
between the possibility to escape the consequences of deleterious
mutations, greater genetic variety, and the number of offspring.
Our findings indicate that the tenet ``better offspring are more important
than the number offspring" might be further
refined; the {\it relative}
improvement of the offspring with respect to the parents turns out to be
a factor in determining the ratio of sexuals to asexuals in the steady
state, and  this dependence is the stronger, the fewer the offspring.

\bigskip
{\bf Acknowledgements}
We are grateful for   useful discussions we have had
with Benan Din\c ct\"urk and Dietrich Stauffer. NJ acknowledges
the hospitality of the G\"ursey Institute and the
Istanbul Technical University, where this research was initiated.
AE would like to thank the Turkish Academy of Sciences for partial support.

\end{multicols}

%%%%%%%%%% TABLES %%%%%%%%%%%%

\newpage
\normalsize
\bc TABLES \ec

\begin{table}
\footnotesize
\bc
\begin{tabular}{|c|c|c|c|c|c|} 
{\boldmath ~~~m~~~} &0 &1&2&3&4\\ \hline
{\boldmath ~~~n(m)~~~}&  1.21  &  9.87 & 35.63  & 53.16 &  0.13 \\
\end{tabular}
\ec
\caption{Distribution with respect to the number of mutations $m$ 
in the purely asexual steady state population in Model I. }
\end{table}

\vskip 2cm

\normalsize
\begin{table}
\footnotesize
\bc
\begin{tabular}{|c|cc|cc|cc|cc|cc|} 
{\boldmath ~m~}& n(m)& (Model I) & n(m) & (Model II) & n(m) &
(Model III) & n(m) & (Model IV) & n(m) & (Model V) \\
  & Asexual & Sexual & Asexual & Sexual & Asexual & Sexual &
Asexual & Sexual & Asexual & Sexual \\ \hline \hline
0 &    0.00 &  1.76  &  0.54   & 17.78  &  0.51   & 17.81
&  1.15   & 2.97  & 1.04 & 4.10 \\
1 &    0.00 & 12.04  &  3.90   & 19.97  &  4.17   & 17.98
&  8.75   & 4.63 & 8.29 & 6.35\\
2 &    0.00 & 35.62  & 13.36   & 14.99  & 14.41   & 14.29  &
30.43   & 4.54 & 28.20 & 5.76\\
3 &    0.00 & 50.53  & 20.03   &  9.39  & 22.05   &  8.74
& 44.69   & 2.75 & 42.56 & 3.64 \\
4 &    0.00 &  0.06  &  0.00   &  0.04  &  0.00   &  0.04
&  0.00   & 0.10 & 0.00 & 0.07 \\ \hline
$\Sigma$ & 0.00& 100.00 & 37.83 & 62.17 & 41.14 & 58.86& 84.93& 15.07&
80.09 & 19.91 \\
\end{tabular}
\ec
\caption{Distribution of the coexisting asexual and sexual steady 
state populations with respect to the number of mutations $m$, for Models 
I-V, after sex has been turned on. $\Sigma$ indicates the 
total percentages of asexual and sexual individuals. }
\end{table}

%%%%%%%%%% FIGURES %%%%%%%%%%%%

\newpage 
\normalsize 
\bc FIGURES \ec 

\begin{multicols}{2}
%Figure 1
\begin{figure}
\begin{center}
\leavevmode
\psfig{figure=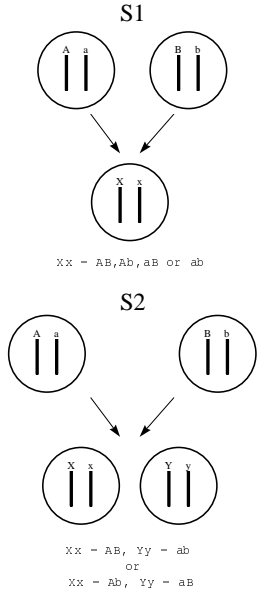,width=4cm,height=8cm,angle=0}
\end{center}
\narrowtext
\caption{ We illustrate how the two bit strings are shared between 
two sexual individuals as they beget one (two) offspring, according to 
the rules {\bf S1} ({\bf S2}) of sexual reproduction.  See text. } 
\end{figure}

%Figure 2 
\begin{figure} 
\begin{center} 
\leavevmode 
\psfig{figure=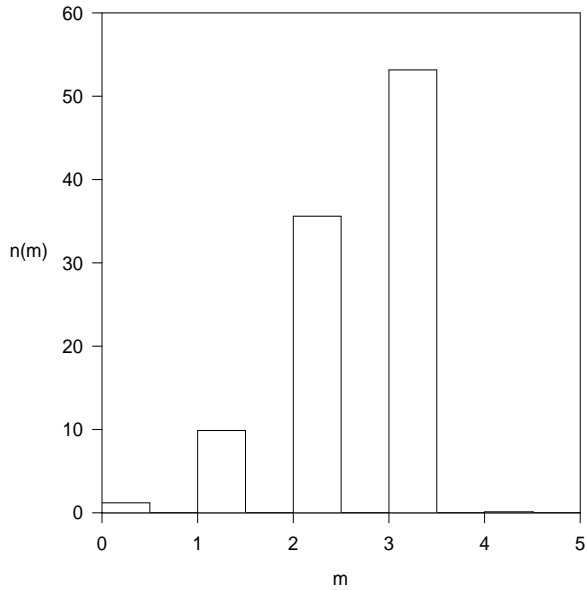,width=8cm,height=8cm,angle=0} 
\end{center}
\narrowtext 
\caption{ The steady state distribution (in percentages) 
of the asexual population after 5000 generations, with respect to the number 
of mutations $m$, before sex is introduced. }
\end{figure} 
\vskip 4.85cm 

%Figure 3 
\begin{figure} 
\begin{center} 
\leavevmode 
\psfig{figure=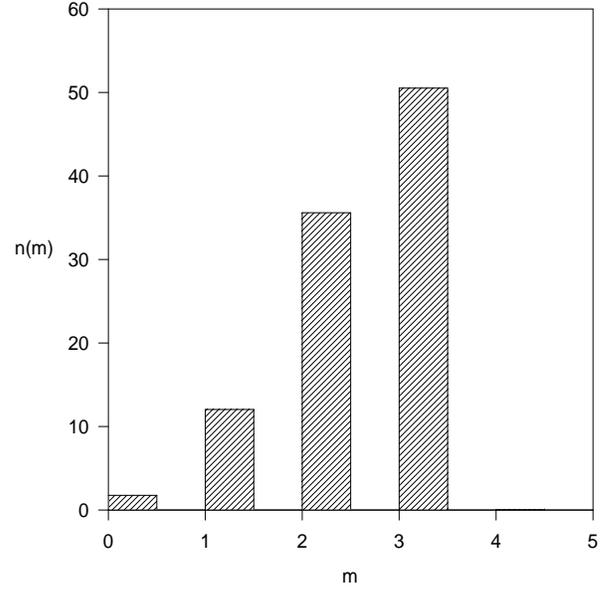,width=8cm,height=8cm,angle=0} 
\end{center} 
\narrowtext 
\caption{ The distribution of the steady state sexual population 
in Model I; $m$ is the number of mutations. }
\end{figure}

%Figure 4
\begin{figure}
\begin{center}
\leavevmode
\psfig{figure=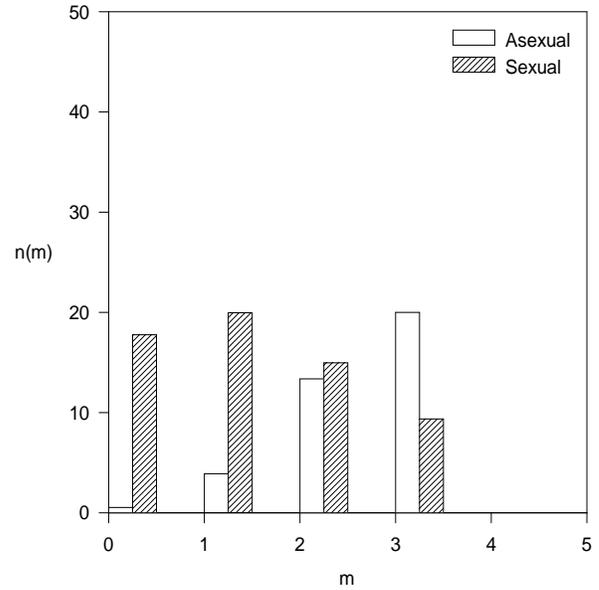,width=8cm,height=8cm,angle=0}
\end{center}
\narrowtext 
\caption{ When sexual individuals are only allowed to reproduce sexually, 
one finds that they reach a finite fraction of the total population, 
and coexist with the asexuals. Here we show the steady state 
distribution (in percentages) of the sexual and asexual 
populations in this coexisting state for Model II. }
\end{figure}

\newpage

%Figure 5
\begin{figure}
\begin{center}
\leavevmode
\psfig{figure=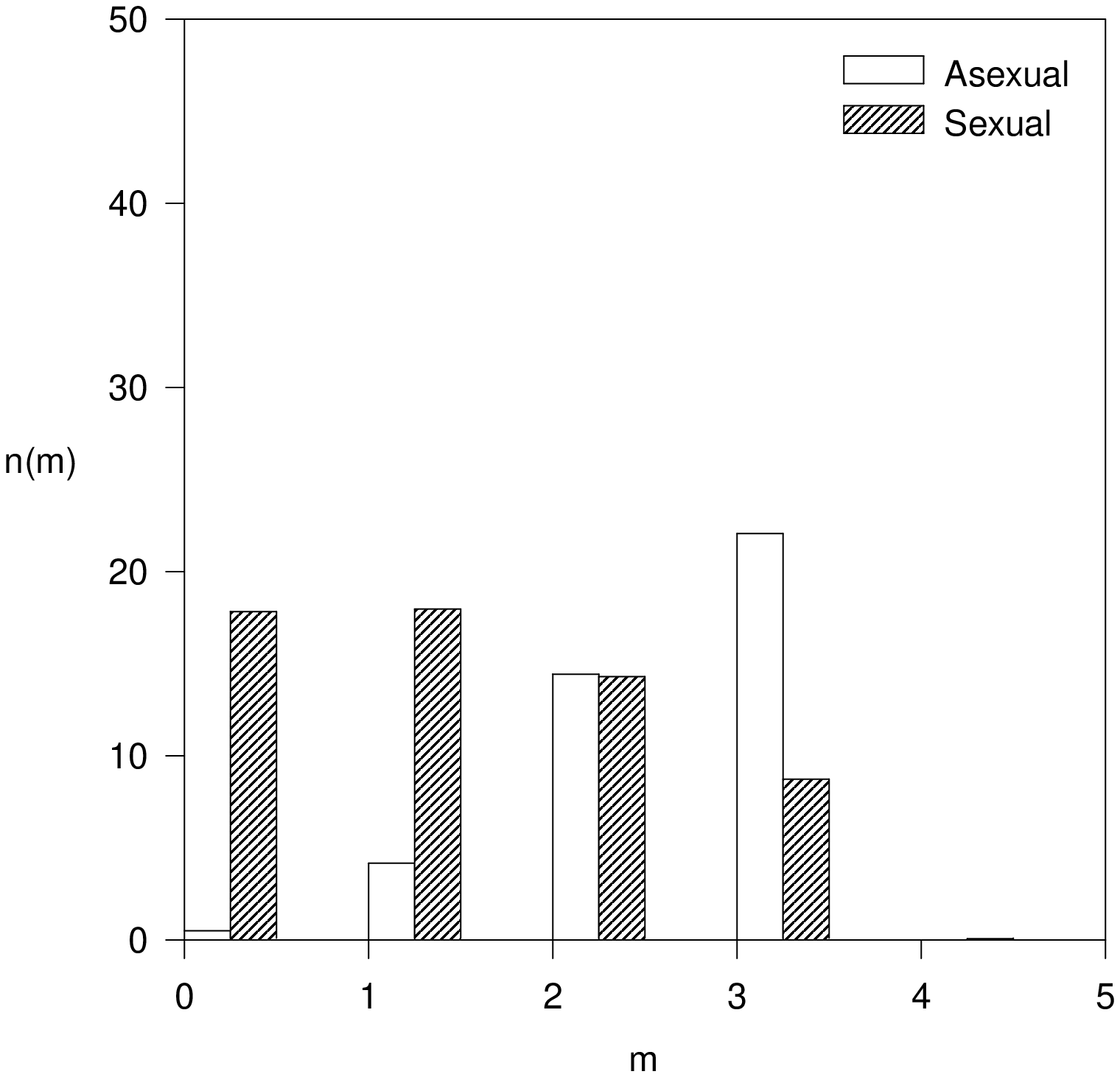,width=8cm,height=8cm,angle=0}
\end{center}
\narrowtext 
\caption{ Steady state distribution of the asexual and sexual 
populations (in percentages) for Model III. }
\end{figure}
\vskip 1cm

%Figure 6
\begin{figure}
\begin{center}
\leavevmode
\psfig{figure=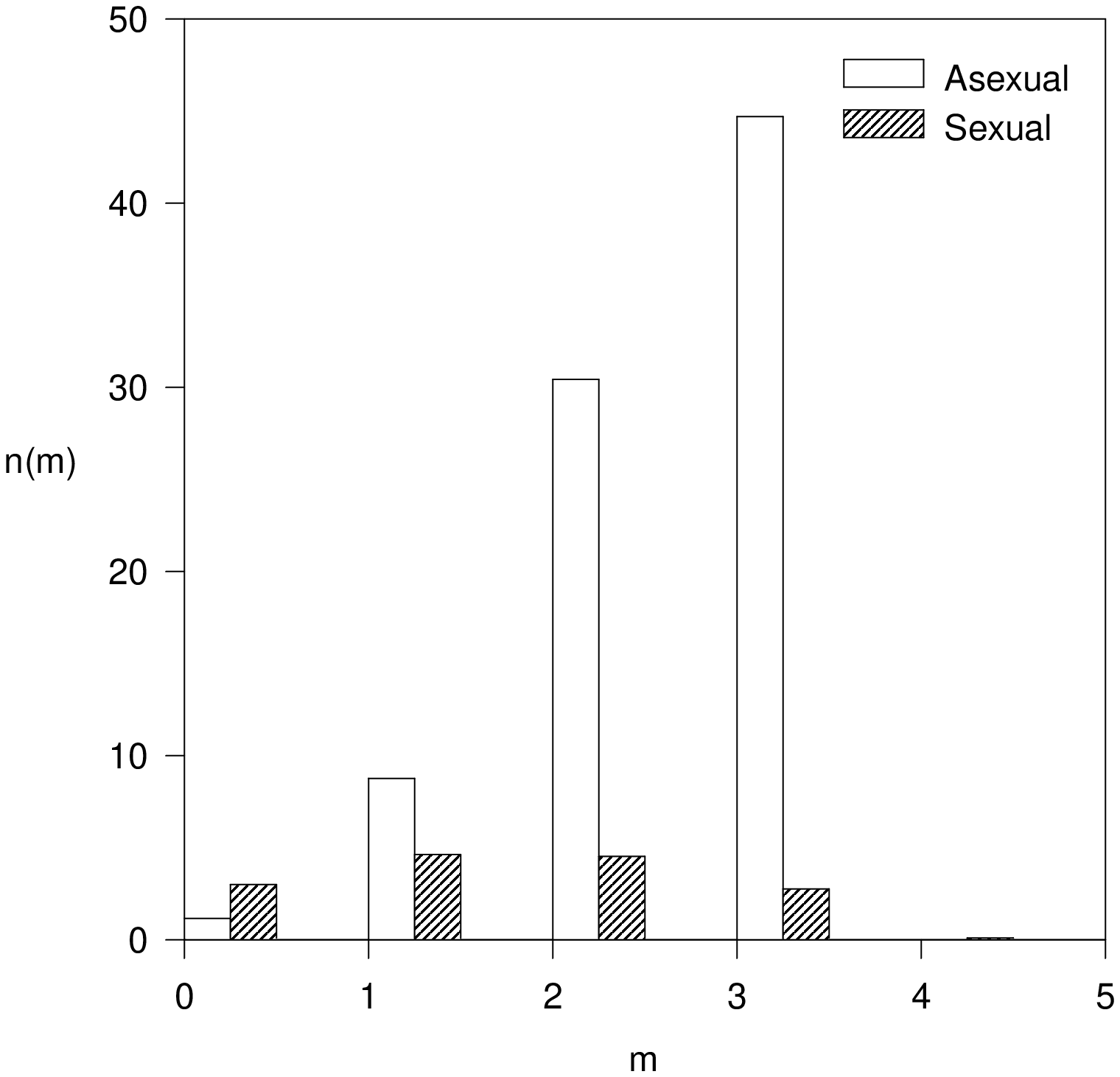,width=8cm,height=8cm,angle=0}
\end{center}
\narrowtext 
\caption{ Steady state distribution of the asexual and sexual 
populations (in percentages) for Model IV. }
\end{figure}

%Figure 7
\begin{figure}
\begin{center}
\leavevmode
\psfig{figure=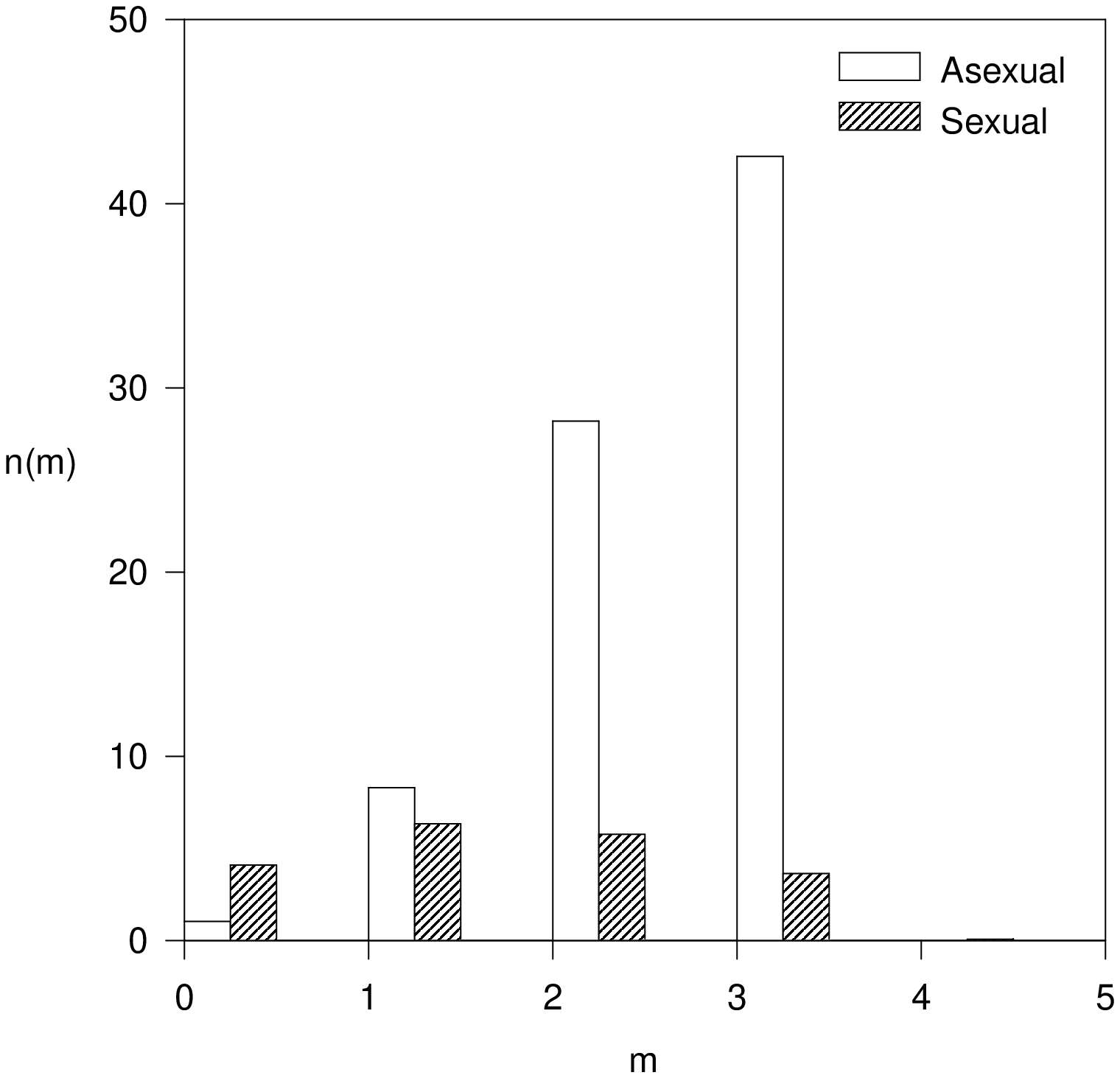,width=8cm,height=8cm,angle=0}
\end{center}
\narrowtext 
\caption{ Steady state distribution of the asexual and sexual 
populations (in percentages) for Model V. }
\end{figure}
\vskip 1cm

%Figure 8
\begin{figure}
\begin{center}
\leavevmode
\psfig{figure=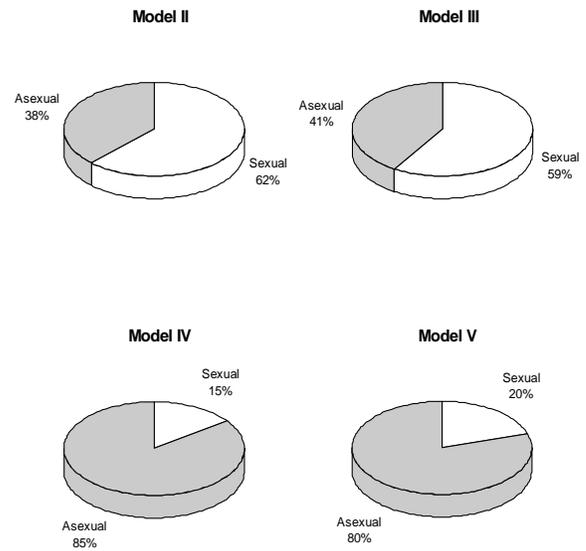,width=8cm,height=8cm,angle=0}
\end{center}
\narrowtext 
\caption{ Pie chart showing the relative weight of the sexual and 
asexual populations in Models II-V. }
\end{figure}

\end{multicols}

\end{document}